%% file: nat-arx2.tex
\documentclass[12pt]{article}
\input macos.tex
\usepackage{graphicx}
\usepackage{graphics}
\usepackage{float}

\def\prd{Phys.Rev.~D}
\def\apjl{Astrophys.J.(Lett.)}

\usepackage[colorlinks,citecolor=black]{hyperref}

\begin{document}

\begin{center} {\large {\bf Multi-Messenger Astrophysics}\footnote{\normalsize Update of arXiv:1906.10212, v1, which appeared in Nature Reviews Physics, 1:585 (2019), https://www.nature.com/articles/s42254-019-0101-z} } \end{center}

\begin{center} 
P\'eter M\'esz\'aros$^{1,2,3,4}$, 
Derek B. Fox$^{1,3,4}$,
Chad Hanna$^{2,1,3,4}$,
Kohta Murase$^{2,1,3,4,5}$ 
\end{center}
{\footnotesize
\noind
$^1$Dept. of Astronomy \& Astrophysics, Pennsylvania State University, University Park, PA 16802, USA\\
\noind
$^2$Dept. of Physics, Pennsylvania State University, University Park, PA 16802, USA\\
\noind
$^3$Center for Particle and Gravitational Astrophysics, Pennsylvania State University, University Park, PA 16802, USA\\
\noind
$^4$Institute for Gravitation and the Cosmos, Pennsylvania State University, University Park, PA 16802, USA\\
\noind
$^5$Center for Gravitational Physics, Yukawa Institute for Theoretical Physics, Kyoto University, Kyoto 606-8502 Japan
}
\normalsize
\\

\noindent
{\bf Abstract}\\
Multi-messenger astrophysics, a long-anticipated extension to
traditional and multiwavelength astronomy, has recently emerged as a
distinct discipline providing unique and valuable insights into
the properties and processes of the physical universe.
These insights arise from the inherently complementary information
carried by photons, gravitational waves, neutrinos, and cosmic rays
about individual cosmic sources and source populations.
{  This complementarity is the basic reason why multi-messenger
astrophysics is much more than just the sum of the parts.}
Realizing the observation of astrophysical sources via non-photonic
messengers has presented enormous challenges, as evidenced by the
fiscal and physical scales of the multi-messenger observatories.
However, the scientific payoff has already been substantial, with
even greater rewards promised in the years ahead.
In this review we survey the current status of multi-messenger
astrophysics, highlighting some exciting recent results, and
addressing the major follow-on questions they have raised. 
Key recent achievements include the measurement of the spectrum of
ultra-high energy cosmic rays out to the highest observable energies;
discovery of the diffuse high energy neutrino background; the first
direct detections of gravitational waves and the use of gravitational
waves to characterize merging black holes and neutron stars in
strong-field gravity; and the identification of the first joint
electromagnetic + gravitational wave and electromagnetic + high-energy
neutrino multi-messenger sources.
We then review the rationales for the next generation of
multi-messenger observatories, and outline a vision of the most likely
future directions for this exciting and rapidly advancing field.
%
%
%
\\
\noindent
\section*{\normalsize Key Points}
\label{sec:keypoints}

\begin{enumerate}

\item Multi-messenger astrophysics aspires to make use of the information 
provided about the astrophysical universe by all four fundamental forces of 
Nature, namely {  the gravitational, the weak and the strong forces,
besides the electromagnetic force which previously had provided almost all our 
information about the Cosmos. These new channels provide previously untapped, 
qualitatively different and complementary types of information, capable of probing 
down to the densest and energy-richest regions of cosmic objects, which were 
hitherto hidden from astronomers' sights}.

\item Diffuse backgrounds of high-energy neutrinos (HENs) with
  energies from $\simg$10\,TeV to PeV, ultra-high energy cosmic
  rays (UHECRs) {  at energies of $\simg 10^{18}\eV$, and }
  $\gamma$-rays with energies between MeV and $\sim$TeV have been
  measured, or upper limits have been provided, by Cherenkov
  detectors, satellites and ground-based air-shower arrays.

\item Gravitational waves (GWs) from merging stellar mass black hole
  and neutron star binaries have been detected at frequencies {  in the
  $\simg$ 10~Hz to $\sim$ 1~kHz range} with laser interferometric
  gravitational-wave detectors.

\item The sources of the diffuse UHECR and HEN backgrounds remain
  unknown, although a gamma-flaring blazar (a type of active galaxy
  with a massive black hole at the center ejecting a relativistic
  plasma jet towards the observer) has been tentatively identified
  with observed HENs.  While up to $\sim$85\% of the $\gamma$-ray
  background can be attributed to blazars, it appears that at most
  30\% of the HEN background can be due to blazars.

\item Formation channels for the observed stellar mass black hole
  binaries, and their possible role as a cosmologically relevant
  component of the dark matter, is currently under debate.

\item There is a natural physical connection between high energy
  cosmic ray interactions and the resulting very high energy neutrinos
  and \grays, which needs to be fully exploited to better understand
  the nature of their unknown astrophysical sources. The connection
  with gravitational wave emission, while less direct, can be expected
  to provide important information about supermassive black hole
  populations and dynamics. 

\item Even before the arrival of the next generation of gravitational
  wave, neutrino, and cosmic ray detectors, the present advanced
  LIGO/VIRGO detectors will be able to detect hundreds of binary
  mergers up to $\sim$Gpc distances; yet electromagnetic (EM) counterpart 
  searches rely primarily on the aging space-based facilities \swift\ and 
  \fermi, currently operating well beyond their design lifetimes. These EM
  counterparts have been found mainly in gamma- or \xrays, and there
  is an urgent need for a new generation of EM detectors, also
  extending into other frequencies including the UV, optical/IR, and
  radio.


\end{enumerate}

\section{Introduction}
\label{sec:intro}

Of the four fundamental forces in nature -- the electromagnetic,
gravitational, weak and strong nuclear forces -- until the middle
of the 20th century it was only messengers of the electromagnetic
force, in the form of optical photons, which allowed astronomers to
study the distant universe. Subsequently, advancing technology added
to these radio, infrared, ultra-violet, \xray\ and \gray\ photons. 
Finally, in the last few decades the messengers of the other three forces, 
namely gravitational waves (GWs), neutrinos, and cosmic rays (CRs), 
began to be used in earnest. {  Thus, we are now finally using the 
complete set (as far as known) of forces of Nature, which are revealing 
exciting and hitherto unknown details about the Cosmos and its denizens.} 

These new non-photonic messengers are generally more challenging to
detect and to trace back to their cosmic sources than most
electromagnetic emissions. When detected, they are usually
associated with extremely high mass or high energy density
configurations, e.g.\ the dense core of normal stars, stellar
explosions occurring at the end of the nuclear burning life of massive
stars, the surface neighborhood of extremely compact stellar remnants
such as white dwarfs, neutron stars or black holes, the strong
and fast varying gravitational field near black holes of either
stellar mass or the much more massive ones in the core of galaxies, or
in energetic shocks in high velocity plasmas associated
with such compact astrophysical sources. This association with the
most violent astrophysical phenomena known means that the interpretation 
of multi-messenger observations requires, and can have implications for,
our theories of fundamental physics, including strong-field gravity,
nuclear physics, and particle interactions.

\begin{figure}[ht]
\centerline{\includegraphics[width=6.0in,height=4.0in]{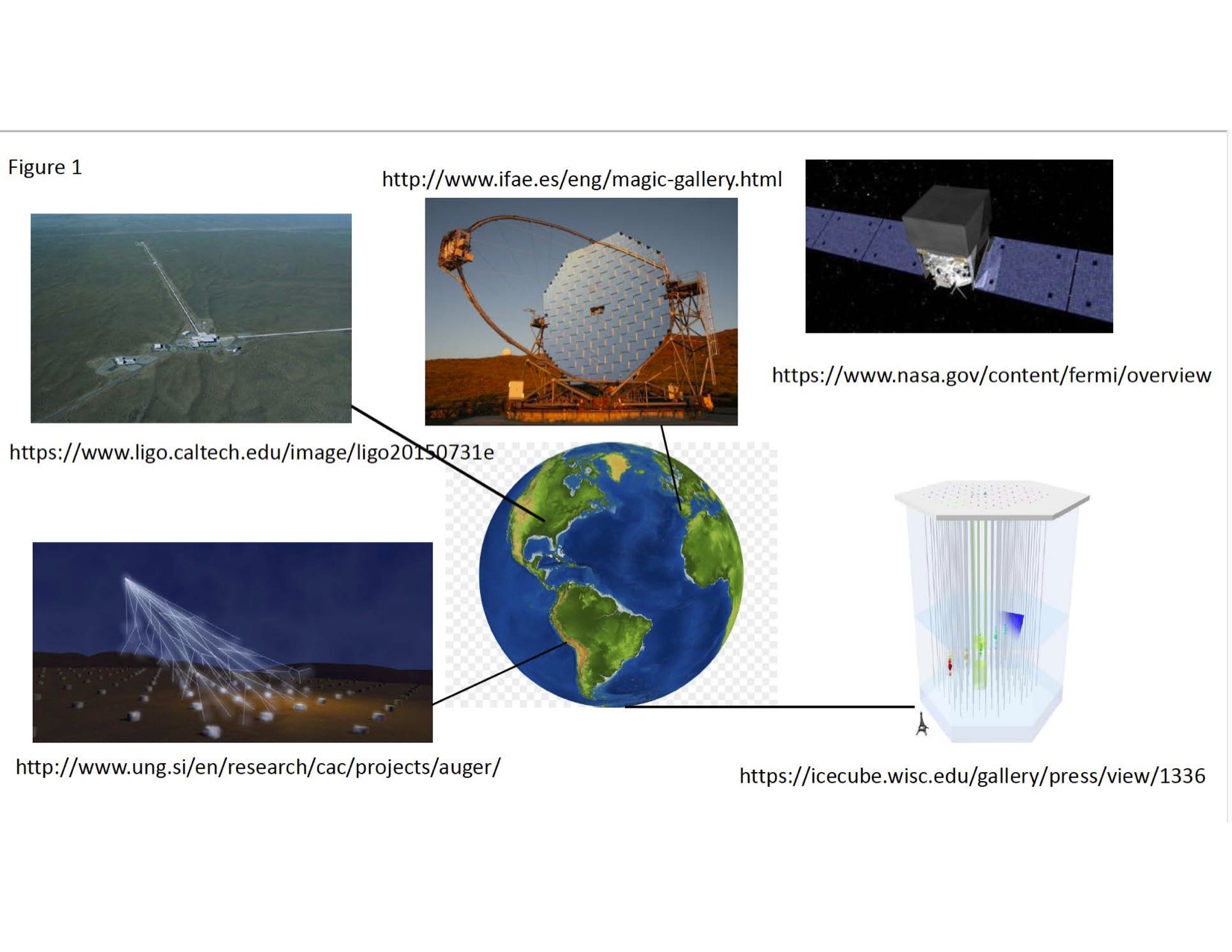}}
\vspace*{-0.5in}
\caption{\footnotesize Examples of current instruments observing
  cosmic messengers via the electromagnetic, weak, gravitational, and
  strong forces, showing  their location. Clockwise from top left: 
the LIGO Hanford gravitational wave interferometer; 
one of the MAGIC air Cherenkov telescopes; 
the Fermi gamma-ray space telescope;
a schemaric of the Pierre Auger cosmic ray observatory in Malarg\"ue, Argentina;
a schematic of the IceCube cubic kilometer neutrino detector in Antarctica; 
}
\label{fig:Fermi-IC3-LIGO-Auger}
\end{figure}

The study of such high energy compact objects started in earnest in
the 1950's, after decades of a slower build-up with increasingly large
ground-based optical telescopes. The first major breakthroughs came
from the deployment of large radio-telescopes, followed by the
launching of satellites equipped with \xray\ and later
\gray\ detectors, which established the existence of active galactic
nuclei, neutron stars, and black holes, and revealed dramatic
high-energy transient phenomena including \xray\ novae, \xray\ bursts,
and gamma-ray bursts (GRBs).
Starting in the late 1960's, large underground neutrino detectors
were built, measuring first the neutrinos produced in the Sun
and later those arising from a supernova explosion; and it was only in
the current decade that extragalactic neutrinos in the TeV-PeV range
were discovered.  Cosmic rays in the GeV energy range started being
measured in the 1910s, but it was only in the 1960s that large
detectors started measuring higher energies implying an extragalactic
origin, and only in the last decade has it become practical to start
investigating the spectrum and composition in the $10^{18}-10^{20}$ eV
range.  Gravitational wave detectors started being built in the 1970s,
but it was not until the 1990s that new technologies and large enough
arrays began to be built approaching the sensitivity required for
detections, the first successes starting in 2015.
For the first time, detectors covering all four fundamental forces of Nature 
{\bf (Fig, \ref{fig:Fermi-IC3-LIGO-Auger})} have been thrown into the breach to explore all 
the previously hidden aspects of the Cosmos.

\section{\Large Mono- and Multi-Messenger Advances} 
\label{sec:currentmm}

The exciting experimental results listed in this section confirm many of the
theoretical/phenomenological predictions and expectations that had been formulated 
over the past several decades, while also bringing up new surprises, which we discuss
in the subsequent section.

\subsection{Recent Non-Photonic Mono-Messenger Results}
\label{sec:singlemes}

\nobu
{\it Ultra-High Energy Cosmic Rays.-}
The Pierre Auger cosmic ray observatory (PAO) \cite{Auger+15revinstr}, located in Argentina, is a 
3,000 km$^2$ array of 1660 water Cherenkov stations and 27 air-fluorescence telescopes (one of the 
tanks and a set of fluorescence telescopes is in {\bf Fig. \ref{fig:Fermi-IC3-LIGO-Auger}, lower right)}, 
designed to detect UHECRs at energies between $10^{17}\eV$ to $10^{21}\eV$.
Its measurements of the diffuse UHECR flux energy spectrum, starting in 2009, confirmed conclusively
the existence of a spectral steepening setting in near $6\times 10^{19}\eV$, e.g. 
\cite{Auger+17icrc17highlight}, which had been first observed by the HiRes instrument 
\cite{Abbasi07hires}.{  This is consistent with the so-called GZK (Greisen, Zatsepin, Kuz'min, 
\cite{Greisen66,Zatsepin+66}) feature expected from CR proton energy losses or from heavy ion
photo-dissociation \cite{Gerasimova+62nucphotodis} due to interactions with cosmic microwave background 
photons, although it could also be due to reaching a maximum acceleration at the sources.}
From 2010 onwards, Auger also started showing evidence for an UHECR chemical composition becoming
heavier above $\simg 10^{18.5}\eV$. The statistical significance of these results has become stronger
over the years \cite{Auger+16mixedcomp,Gora+18augerrev,Petrera+19augerrev}. The spectral results are
consistent, within statistical uncertainties with those obtained with the smaller Telescope Array (TA)
UHECR observatory \cite{Kawai+08-TA}, which is important because Auger is in the Southern hemisphere while 
TA is in the Northern. {  The chemical composition issue \cite{TA+18masscomp} is still under debate,
although a joint Auger-TA paper \cite{AbuZayyad+18augerTAchem} shows results which agree within the errors.}
The angular resolution in the arrival direction of UHECRs is below $1^o$ above $\sim 10^{19}\eV$ for
both protons and heavy elements, although the magnetic deflection increases with mass;
around $10^{19}\eV$ it is $\siml 5^o$ for protons, while for heavy nuclei it could be tens of degrees.
At these energies, due to 
the energy losses caused by the GZK effect mentioned, these UHECRs must have originated within
distances of $\siml 100\Mpc$.  So far, all attempts at finding angular spatial correlations between 
UHECRs and any type of known cosmic sources have been unsuccessful \cite{Auger+17icrc17highlight}.
\\

\nobu
{\it High Energy Neutrinos.-}
The IceCube neutrino observatory \cite{Achterberg+06-ICECUBE} consists of a cubic kilometer (roughly a 
Gigaton) of ice at a depth between 1.4 and 2.4 km below the South Pole, instrumented with 86 strings 
connecting 5,160 optical phototubes (see schematic in {\bf Fig. \ref{fig:Fermi-IC3-LIGO-Auger}, 
lower right)}, which measure the light radiated from {  charged particles} produced by passing high energy 
neutrinos interacting with the ice. Its construction was finished in 2010, and in 2012-2013 it discovered
a diffuse flux of neutrinos in the range $100 \TeV \siml E_\nu \siml` 1 \PeV$ \cite{IC3+13pevnu1,
IC3+13pevnu2}, later extended down to $\siml 100\TeV$. The energy spectrum $dN/dE_\nu$ can be fitted 
with a $\sim -2.5$ index power law, but there may be an indication for  two components, steeper below $\sim
200\TeV$ and flatter (index $\sim -2$ above that, the highest energy so far being $\sim 10\PeV$.
IceCube detects all neutrino flavors, with muon neutrino charged current interactions resulting in 
elongated Cherenkov tracks and all other neutrino flavors and interactions largely producing 
{near-spherical optical Cherenkov signals from secondary particle cascades,}
the direction of arrival being uncertain by $\sim 10^o -15^o$ for cascades 
and $\sim 0.5^o-1.0^o$ for tracks. 
Tau neutrinos can also be identified at sufficiently higher energies, where the statistics are lower, 
and these have  not yet been identified, 
{  although a suggested tau-like candidate has been discussed \cite{Kistler+18multipev}.}
The observed flavor distribution is compatible with complete flavor mixing having occurred due to
the neutrino oscillation phenomenon over cosmological distances \cite{Halzen17nurevnat,Ahlers+18ic3nurev}.
So far there is no evident correlation of the observed neutrinos with any type of known cosmic 
objects \cite{IC3+17icrc17nusources}, except for one interesting case discussed below.
{The smaller underwater Cherenkov telescopes ANTARES \cite{ANTARES+19results} and Baikal-GVD
\cite{Baikal+18status} have also been in operation and providing upper limits.}
At much higher energies, the high altitude balloon experiment ANITA \cite{Allison+18anita3res}, 
flying in a circumpolar orbits in Antarctica, has used a radio
technique to measure neutrinos at $\simg 10^{17}$ eV, which is starting to provide constraints on
cosmological neutrino sources and the GZK-related cosmogenic neutrino fluxes, complementary with
those provided by Auger \cite{Auger+15cosmonu}.
\\

\nobu
{\it Gravitational Waves.- }
The Laser Interferometric Gravitational Wave Observatory (LIGO) consist of two detectors, in Louisiana 
and Washington state, each with 4 km long L-shaped arms, 
{\bf (Fig. \ref{fig:Fermi-IC3-LIGO-Auger}, upper left)} which in 2015 began operation in the 
$\sim 10 - 10^3$ Hz frequency range \cite{Abbott+09-LIGO}.  
Another array, VIRGO \cite{Acernese+15-VIRGO}, located near Pisa, Italy, and similarly L-shaped with 
3 Km long arms, has been operating at epochs coincident with LIGO. Both are actively being commissioned 
and will achieve design sensitivity in the coming years.
The long-awaited first discovery of gravitational waves from a 
stellar mass binary black hole merger (labeled GW150914) was announced by LIGO in 2016 
\cite{LIGO+16-gw150914disc} {\bf (Fig. \ref{fig:BBHGW-crnugam-BNS-blazarnu}(a)}. 
This was soon followed by a number of other binary black hole (BBH) 
mergers detected both by LIGO and, with lower statistical significance, by VIRGO as well 
\cite{LIGO+18bbhGWTC-1}. These BHs weigh up to several tens of solar masses, and have low spins.
%
%
\noindent
However, despite intensive searches, no other messengers associated with BBH mergers have been detected
so far, except for a possible $\gamma$-ray burst \cite{Connaughton+16-gbmgw150914} in GW150914.
\\

\nobu
{\it Electromagnetic Detections.- }
Except for the binary black holes, all the other sources detected with other messengers had been 
previously extensively studied through their EM emissions at various wavelengths. 
Of major recent relevance are the observations in the optical, X-ray and up to 150 keV  
$\gamma$-rays with the Swift satellite, and between 10 keV X-rays to $\siml \TeV$ $\gamma$-rays 
with the Fermi satellite {\bf (Fig. \ref{fig:Fermi-IC3-LIGO-Auger}, upper right)}
\cite{Thompson15spacegam}, which detected a large number of 
Gamma-Ray Burst (GRB) sources, active galactic nuclei (AGNs) including blazars, supernovae, etc., 
as well as a diffuse cosmic $\gamma$-background.  
Of increasing importance for such sources are the ground-based air Cherenkov imaging telescopes,
e.g. MAGIC {\bf (Fig. \ref{fig:Fermi-IC3-LIGO-Auger}, upper middle)}, HESS, VERITAS 
\cite{Montaruli19ctalhaasogam} and the High Altitude Water Cherenkov observatory HAWC 
\cite{HAWC-instr18,Casanova+18hawcrev}, which measure gamma-rays in the 100 GeV to multi-TeV range. 
These have been amply supported by ground and space observations with multiple radio, infra-red, optical 
and UV telescopes.  

\subsection{Developments in Joint Multi-Messenger Astrophysics}
\label{sec:recentmultimes}

\nobu
{\it Solar and Supernova Neutrinos and Photons.- }
The two earliest multi-messenger detections involved neutrinos in the MeV range. Davis and collaborators, 
starting in the 1960's, detected the electron neutrinos produced by the nuclear reactions that are 
the energy source for the Sun's light, using a 600 ton perchlorethylene (cleaning fluid) tank
located deep underground in the Homestake mine in South Dakota, US. This neutrino flux was 
confirmed by various other experiments including the one in the Kamioka mine in Japan, by Koshiba and 
collaborators.
The  other early multi-messenger detection was that of neutrinos from a core-collapse supernova, 
SN 1987a, resulting from inverse beta decay as protons are converted into neutrons. This was
detected by three different underground detectors, Kamiokande in Japan, Baksan  in the Soviet Union, 
and Irvine-Michigan-Brookhaven in  the US
\cite{Hirata+87nusn1987a-Kamioka,Alexeyev+88nusn1987a-Baksan,Haines+88nusn1987a-IMB}.
The neutrino detection preceded significantly the spectacular optical brightening characterizing 
supernovae.  These discoveries earned Davis and Koshiba the Physics Nobel Prize in 2002 
\cite{Davis03Nobel,Koshiba03Nobel}.
\\
\begin{figure}[ht]
\begin{minipage}[t]{0.43\textwidth}
\centerline{\includegraphics[width=3.0in,height=1.5in]{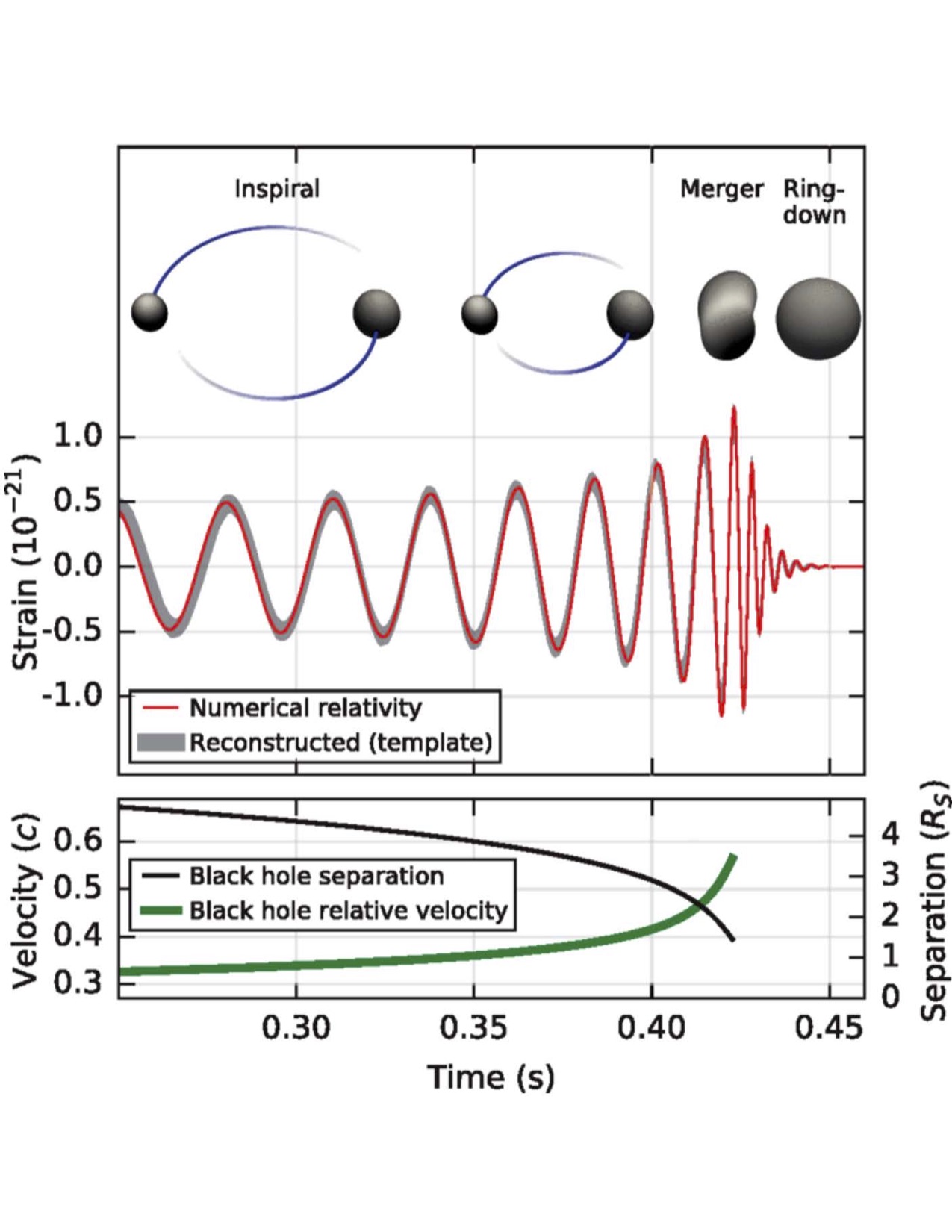}}
\end{minipage}f
\hspace{-40mm}
\begin{minipage}[t]{0.43\textwidth}
\centerline{\includegraphics[width=3.0in,height=1.5in]{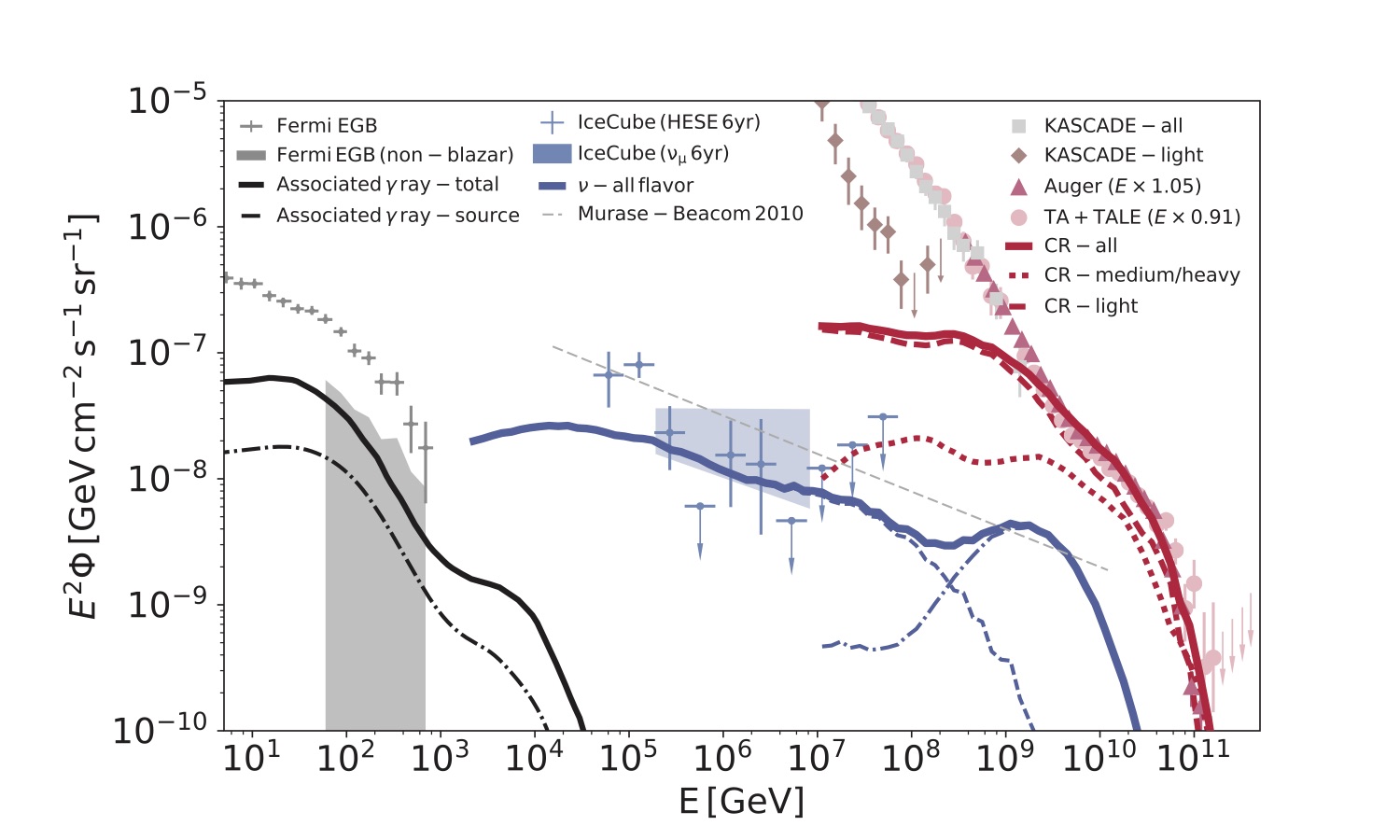}}
\end{minipage}
\hspace{3mm}
\begin{minipage}[t]{0.43\textwidth}
\centerline{\includegraphics[width=3.0in,height=1.5in]{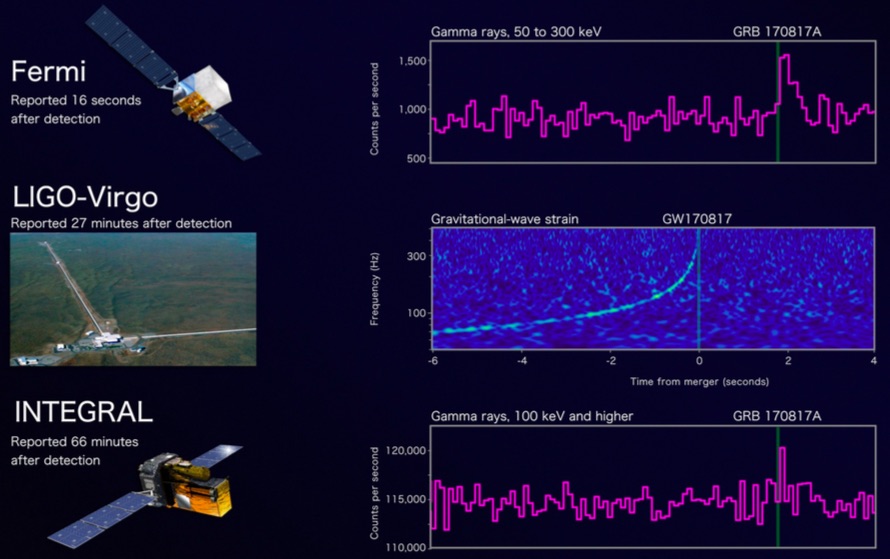}}
\end{minipage}
\hspace{5mm}
\vspace*{-1.1in}
\begin{minipage}[t]{0.43\textwidth}
\centerline{\includegraphics[width=3.0in,height=1.5in]{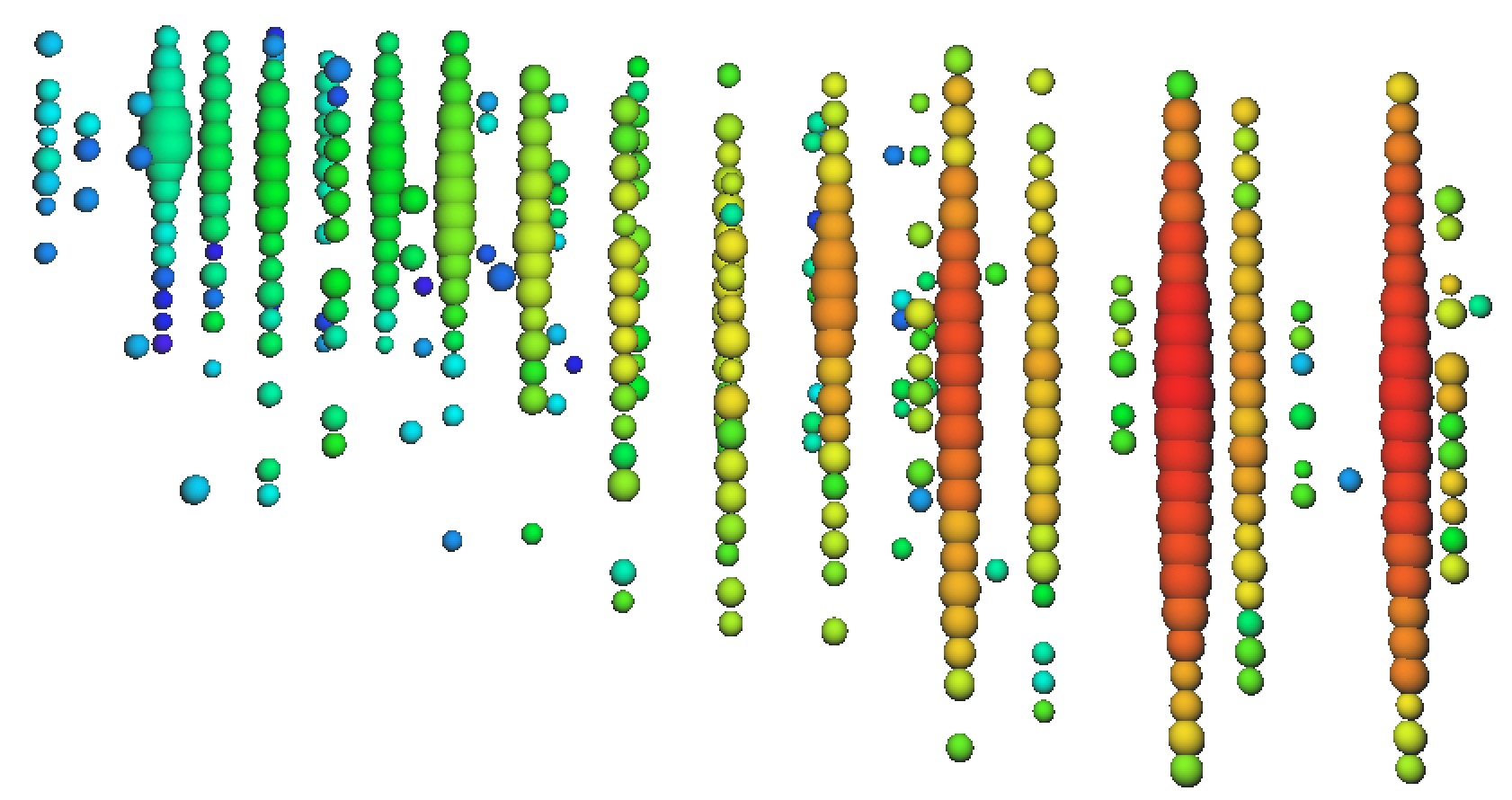}}
\end{minipage}
\vspace*{1.0in}
\caption{\footnotesize 
Some recent cosmic multi-messengers advances involving the electromagnetic,
weak, gravitational and strong forces.
(a) The first GW detection from LIGO/VIRGO in the O1-O2 observing run, of the GW150914 binary black hole 
merger \cite{LIGO+16-gw150914disc}, showing for the inspiral, merger and ringdown phases the theoretical 
and measured waveform, the separation and the relative velocity.
(b) Interrelation expected between (from left to right) the energy spectrum of the diffuse backgrounds
in gamma-rays, high energy neutrinos and ultra-high energy cosmic rays, based on a black hole jet
source model \cite{Fang+18crnuagnjet};
(c) LIGO, VIRGO and Fermi simultaneous multi-messenger discovery of the binary neutron
star merger GW/GRB 170817;
(d) Light track of the muon produced by a 290 TeV muon neutrino coming from  the direction of the
blazar TXS 0506+056, detected on 22 September 2017 by IceCube.
  }
\label{fig:BBHGW-crnugam-BNS-blazarnu}
\end{figure}
%

\nobu
{\it Cosmic ray, Gamma-ray and Neutrino Background Interdependences.- }
The measurements of the diffuse UHECR energy spectrum by the Pierre Auger Observatory starting in 2008
put on a firm ground the detection of a spectral cutoff above $10^{19.5}$ eV, compatible with the 
GZK energy losses due to the cosmic microwave background photons \cite{Auger+17icrc17highlight},
after earlier work by HiRes.
Then, starting in 2008, the Fermi satellite (following on previous work by COS-B and other missions) 
measured a diffuse gamma-ray background extending into the sub-TeV range \cite{Ackermann+15igbfermispec}.
And starting in 2012-2013 IceCube discovered, with increasing detail, a diffuse high energy neutrino (HEN)
background of astrophysical origin at multi-TeV to PeV energies \cite{IC3+13pevnu1,IC3+13pevnu2}.
There is so far no firm identification of the sources of either the UHECR, HEN or gamma-ray diffuse 
backgrounds, although the extragalactic gamma-ray background is known to be dominated by blazars
\cite{Ackermann+16igbfermi}. However, theoretical relationships and mutual constraints are expected 
from the basic physics of these three radiations, e.g. {\bf Fig. \ref{fig:BBHGW-crnugam-BNS-blazarnu} (b)}. 
%
%
\noindent
The HENs are produced when UHECRs collide with low energy 
target photons and nuclei resulting in charged and neutral pions, which decay in a predictable 
fraction of high energy neutrinos and gamma rays.  The resulting energy spectra of neutrinos 
and photons imply corresponding diffuse backgrounds which must fit the observed results, 
including also the constraint provided by the observed UHECR background. 
The fact that the energetics of these three messengers is comparable has led to the idea of a 
unification of high-energy cosmic particles, e.g., \cite{Murase+16uhenurev,Fang+18crnuagnjet}.
On the other hand, significant constraints are also placed on generic $pp$ hadronuclear production 
models of HENs and gamma-rays when when one compares them to the Fermi diffuse $\gamma$-ray flux, 
especially accounting for the $\gamma\gamma$ cascades { initiated by $\gamma$-rays scattering
off cosmic radiation backgrounds} \cite{Murase+13pev,Murase+16hidden}.
The constraints are more stringent for Galactic sources \cite{Ahlers+14pevnugal}.
HAWC \cite{HAWC-instr18} is expected to uniquely contribute to measurements of the $\gamma$-ray background 
in the 10 to 100 TeV energy range, which could strongly constrain the fraction of IceCube neutrinos from 
Galactic  origin.
Among $p\gamma$ photomeson production models of HENs, valuable constraints have been put on the 
contribution of the simpler classical GRB neutrino emission models \cite{Abbasi+12-IC3grbnu-nat,
IC3+15promptnugrb,IC3+18grbnuconstraint}, while leaving open the possibility of contribution by 
choked GRBs or supernovae driven by choked jets \cite{Meszaros+01choked,Murase+13choked,
IC3+18grbnuconstraint}.
\\

%
\nobu
{\it Gravitational Waves and Photons from Binary Neutron Star (BNS) Mergers.- }
{  As the culmination of a long series of previous BNS GW/multi-messenger searches, e.g.  
\cite{LIGO+05grb030329,LIGO-VIRGO+08gwcounter,Kanner+08gwopt,LIGO+08grb070201}, the}
joint GW/EM detection of the transient GW/GRB 170817 was the first high significance proof of
the strength of the joint multi-messenger technique in the GW realm \cite{LIGO+17gw170817disc},
e.g. see {\bf Fig. \ref{fig:BBHGW-crnugam-BNS-blazarnu} (c)}.
%
%
\noindent
The GWs in GW/GRB 170817 showed that this was a neutron star binary merger, providing a measurement 
of their masses, the distance \cite{LIGO+18-170817properties} and gave constraints on the neutron star 
equation of state \cite{LIGO+18-170817nsradeos}, while $\gamma$-ray and X-ray measurements by Fermi 
and Swift showed it was an off-axis short GRB, e.g.  \cite{Troja+17gw170817xr}.
The near simultaneous observation of EM and GW signals from GW170817 showed that they both travel at the 
speed of light to better than 1 part in $10^{15}$, thereby ruling out many alternative theories of gravity.  
Optical observations with various telescopes showed that it also manifested itself as a Kilonova, 
which is an outflow rich in the so-called r-process high atomic number nuclear elements,
providing a {  rich interlocking picture, e.g. \cite{Coulter+17gw170817,Kasliwal+17gw170817concord,
Abbott+17-170817multi,Margutti+17-170817xr}. }
{  For their role in the discovery of binary gravitational wave sources Barish, Thorne and Weiss
received the 2017 Nobel Prize in Physics \cite{Weiss19Nobel,Barish19Nobel,Thorne19Nobel}.}
\\

\nobu
{\it High Energy Neutrinos and Gamma-rays from Blazars.-} 
The joint neutrino \cite{IC3multi+18txs0506} and electromagnetic detection 
\cite{Magic+17blazarnu,Keivani+18txs0506,NuSTAR+17blazarnu} of the flaring blazar TXS 0506+056 was an 
extremely exciting result, being the first time that a known source was shown to be associated
{  (albeit at the $\sim 3\sigma$ level)} with a high significance astrophysical high energy neutrino
{\bf (Fig. \ref{fig:BBHGW-crnugam-BNS-blazarnu}(d)}.
%
%
\noindent
Blazars are active galactic nuclei (AGNs), which are galaxies with a massive central black hole 
powering a relativistic jet outflow pointing close to the observer line of sight; they are classified 
into BL Lac objects and flat spectrum radio-quasars (FSRQs), TXS 0506+056 appearing to be of the BL Lac 
type \footnote{Recently, however, arguments have been presented \cite{Padovani+19txs0506fsrq} 
indicating that TXS 0506+056 may be an FSRQ instead of a BL Lac object as thought previously.}.
Blazars are notorious for exhibiting sporadic and intense gamma-ray flaring episodes, one of which
was in progress at the time the track-type neutrino was observed. Further analysis indicated that
in previous years other neutrinos may have been associated with this source \cite{IC3+18txs0506a}.
This provided valuable constraints on the radiation mechanisms and the sources of the diffuse 
HEN background. Based on simple one-zone emission models where both HENs and gamma-rays originate
from the same region, the neutrino is a low probability event \cite{Gao+19txs0506,Keivani+18txs0506,
Cerruti+19txs0506} and based on a stacking analysis of HENs and blazars it appears that the blazar 
population as a whole may account for $\siml 10-30\%$ of the entire IceCube neutrino background
{ \cite{IC3+17fermiblaznustack}},
so other sources may in any case need to be appealed to.

\section{\large Emerging Questions and Challenges}
\label{sec:newquations}


\nobu {\it The Lack of EM or HEN counterparts of binary black hole mergers} is
frustrating, with ten binary black hole mergers detected in GWs so far
(as of March 2019). Such emissions are expected to be faint at best
in BBHs, e.g. \cite{Perna+16gw150914grb,Murase+16bbh}, but they would
be very useful for a better understanding of the binary origin and environment, 
{  as well as to get a far better localization than provided by the GWs,
e.g. \cite{Murase+16bbhmulti,Bartos+17bbh,Ford+19bbhagn,LIGO+19multisearch}.}
A much larger sample of BBHs will be needed extending to both 
smaller and larger masses to test the hypothesis that BBHs
provide a cosmologically important dark matter component,
e.g. \cite{Bird+16gw150914,Magee+17bhdm,Carr19bhdm}.  \\

\nobu {\it Detection of HEN from GW/EM-detected binary neutron stars}
would provide an example of a ``triple-messenger" source, and would 
clarify major open questions in our understanding of these objects. 
{  Possible signals and observing strategies have been discussed in,
e.g., \cite{Bartos+11multigwnu,Ando+13multinu,LIGO+19multisearch}.}
Expected HEN fluxes are low, especially for off-axis jet
viewing \cite{Kimura+17sgrbnu,Kimura+18transejnu}, but in the best
case they may be marginally detectable by IceCube (or more plausibly, by a
future IceCube-Gen 2), and would greatly aid in clarifying the physics
of the relativistic jet and the larger-angle slower outflows which give
rise to the GRB, the afterglow, and the kilonova emission of these events.  \\

\nobu {\it Confirmation or refutation of the occurrence of HEN flares in blazars}
{ through additional observations of TXS 0506+056 and other AGNs} 
is urgently needed to address the origin of
the IceCube background and illuminate any possible connection between
the HEN and UHECR backgrounds. Progress in these studies will also
require more targeted calculation of AGN neutrino production models,
yielding detailed predictions for \xray\ and other EM constraints. The
stacking analyses of blazar EM flares against observed HENs
\cite{IC3+17fermiblaznustack,Hooper+18nuagn} as well as theoretical
arguments \cite{Murase+18txs0506} indicate that sources other than
blazars must provide the dominant contribution to the HEN background,
and observational correlation studies involving alternative source
candidates may need to be undertaken, e.g. \cite{Murase+16uhenurev}.  \\

\nobu {\it The masses and spins of the GW-detected compact mergers} offers
new puzzles. One is the origin of the "heavy" binary stellar black holes 
($\simg$ 30 solar masses), it is not clear how they form and evolve. Another
question is why do the LIGO BHs have very low spins or spins mis-aligned with 
the orbital angular momentum. This is in contrast to X-ray BH candidates, 
some of which have very large spins.  Also the fate of the remnant in the BNS 
merger GW 170817 is unknown, e.g. how long did the remnant last before turning 
into a black hole, if it finally did.  {  These and related questions are 
discussed in, e.g., \cite{LIGO+18bbhGWTC-1,LIGO+18bbho1o2}.}
Future GW observations could resolve this issue. \\

\nobu {\it UHECR arrival direction uncertainties are large}, and UHECR
arrival times are delayed by $\sim$10$^4$ to 10$^5$ years relative to
any simultaneous EM or neutrino emission, so direct correlation
attempts have been made only against quasi-steady, non-bursting
sources, so far unsuccessfully \cite{Auger+17icrc17highlight,
  IC3+16-nucrcorrel,Moharana+15=IC3correlcr}.  At the highest
energies, UHECR positional correlations with muon neutrino tracks, UHE
neutrinos, and/or \grays\ could lead to a better pinpointing of the sources. 
This will require {  much better instruments and} more sensitive neutrino/EM 
correlation analyses as well as much more detailed production models for likely 
source candidates.  \\

\nobu {\it Statistically significant measures {  of UHECR/EM/HEN/GW }
  correlations (or lack thereof)} are urgently needed, and it is also
necessary to explain the UHECR spectrum and chemical composition with
an appropriate distribution of specific sources,
e.g.\ \cite{Aloisio+13comp,Alves+19cosmonu}. Observations must be
fitted in statistical detail to model predictions of possible
candidates, such as AGNs, GRBs, tidal disruption events, clusters of
galaxies, etc., and more sophisticated models must be calculated, and
tested against the observed diffuse neutrino and \gray\ backgrounds. \\

\nobu {\it Theory and simulations are still in their infancy}, as far 
as  UHECR, HEN and GW sources. 
{  While the HEN/EM inter-relation is in principle straightforward, 
aside from non-linearities introduced by EM and  hadronic cascades, an
understanding of the HEN/EM inter-relations with UHECRs is significantly 
complicated by the fact that the latter are charged, and hence travel in 
complicated paths, dependent on the magnetic fields, e.g. \cite{Murase+19crnurevfuku}; 
in addition, at energies above $\simg 10^{17}-10^{18}\eV$, cross-section
uncertainties start to set in.}
For BBH and BNS mergers, a lot of progress 
is urgently required to understand post-merger dynamics, the final state 
of the remnant, the physics of the ejecta and how BH-NS mergers differ 
{  from BNS mergers \cite{Radice+18bnsmerg}. 
Supernova simulations have also been a challenge \cite{Glas+19sn3dsimul,Radice+19sncollgw}. }
Lack of reliable GW waveforms means that we have to rely on sub-optimal techniques 
for their detection, and it also makes it far more difficult to distinguish 
between different collapse models/scenarios.\\

\nobu {\it The ultra-high energy neutrino range $10^{17}-10^{20}\eV$
  explorations by ANITA} \cite{Allison+18anita3res} and other future
experiments need to achieve at least an order of magnitude greater
sensitivity to probe the { cosmogenic neutrino background. 
This is due to UHECRs interacting with the cosmic photon backgrounds, 
and degeneracies are induced by the effect of the UHECR source luminosity
function and redshift distribution as well as the CR chemical composition,
e.g. \cite{Stanev+05diffnu,Kotera+11uhecr,Globus+17inbigb}.}
The ANITA anomalous upward-going events \cite{Gorham+18anitataunu}, if
confirmed, are very exciting for what they may tell us about cosmic
tau-neutrinos, or possibly about beyond the standard model
physics. Concordance studies between ANITA, IceCube and Auger will
need to be carried out, together with significantly more detailed
theoretical investigations, e.g. \cite{Alvarez+18taunu,
Romero-Wolf+18anitaup,Connolly+18anitatau,Fox+18anitabsm}.

\section{\large Looking Ahead: New Instruments \& Results Expected}
\label{sec:forward}

The spectacular results achieved, mainly in the last decade, by 
multi-messenger facilities such as those in the first row of Table 1, 
has opened wide new vistas in high energy astrophysics. This has spurred the 
building and planning of more sophisticated and more powerful experiments and 
missions, geared towards the elucidation of the key new questions raised. 
The second row of Table 1 shows some of the new experiments currently 
under constructions, while the third row shows some of the next generation of 
experiments planned for the period between approximately 5 to 15 years from now.
\begin{figure}[ht]
\centerline{\includegraphics[width=6.0in,height=4.0in]{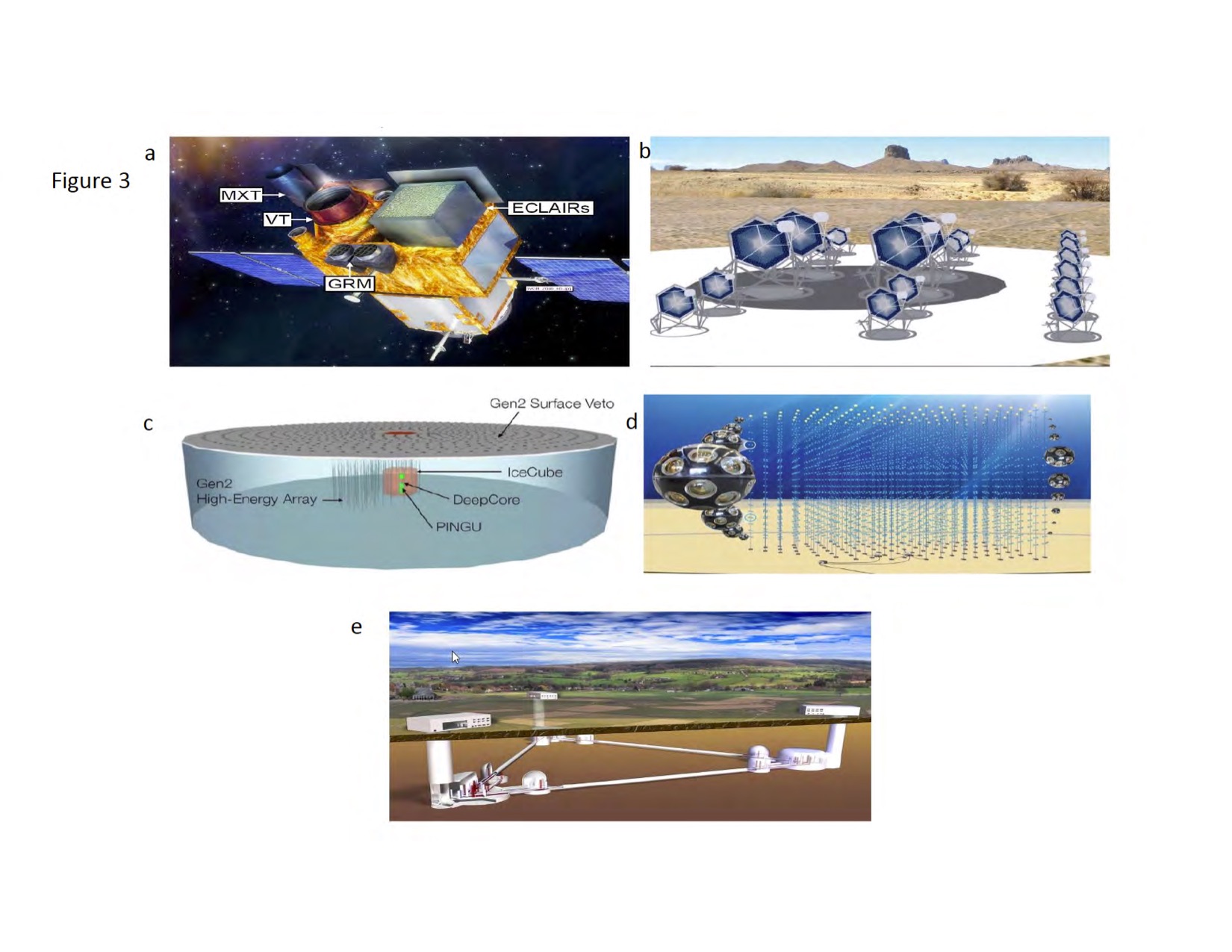}}
\vspace*{-0.5in}
\caption{\footnotesize 
Some major new detectors in the planning stage:
(a) SVOM China-France GRB multi-wavelength follow-up satellite, exp. 2022 \cite{Dagoneau+18svom};
(b) Schematic of the CTA Cherenkov Telescope Array gamma-ray ground array, exp. 2024 \cite{Wild+18-CTA};
(c) IceCube-Gen2, including current IceCube and DeepCore, and the planned high energy array,
super-dense PINGU sub-array and extended surface array (larger ARA radio array not shown) \cite{IC3+17icrcgen2};
(d) KM3NeT planned 3-4 km$^3$ neutrino detector planned in the Mediterranean sea,
which will include also the high-energy ARCA and low energy ORCA sub-arrays \cite{Km3net+18sens};
(e) Schematic of the planned EU next generation Einstein gravitational wave interferometer 
\cite{Sathyaprakash+12-EINSTEIN}.
}
\label{fig:SVOM-CTA-IC3Gen2-KM3NeT-ET}
\end{figure}
\begin{figure}[H]
\vspace*{-0.25in}
\begin{minipage}[t]{\textwidth}
\centerline{\includegraphics[width=8in,height=6in,angle=90]{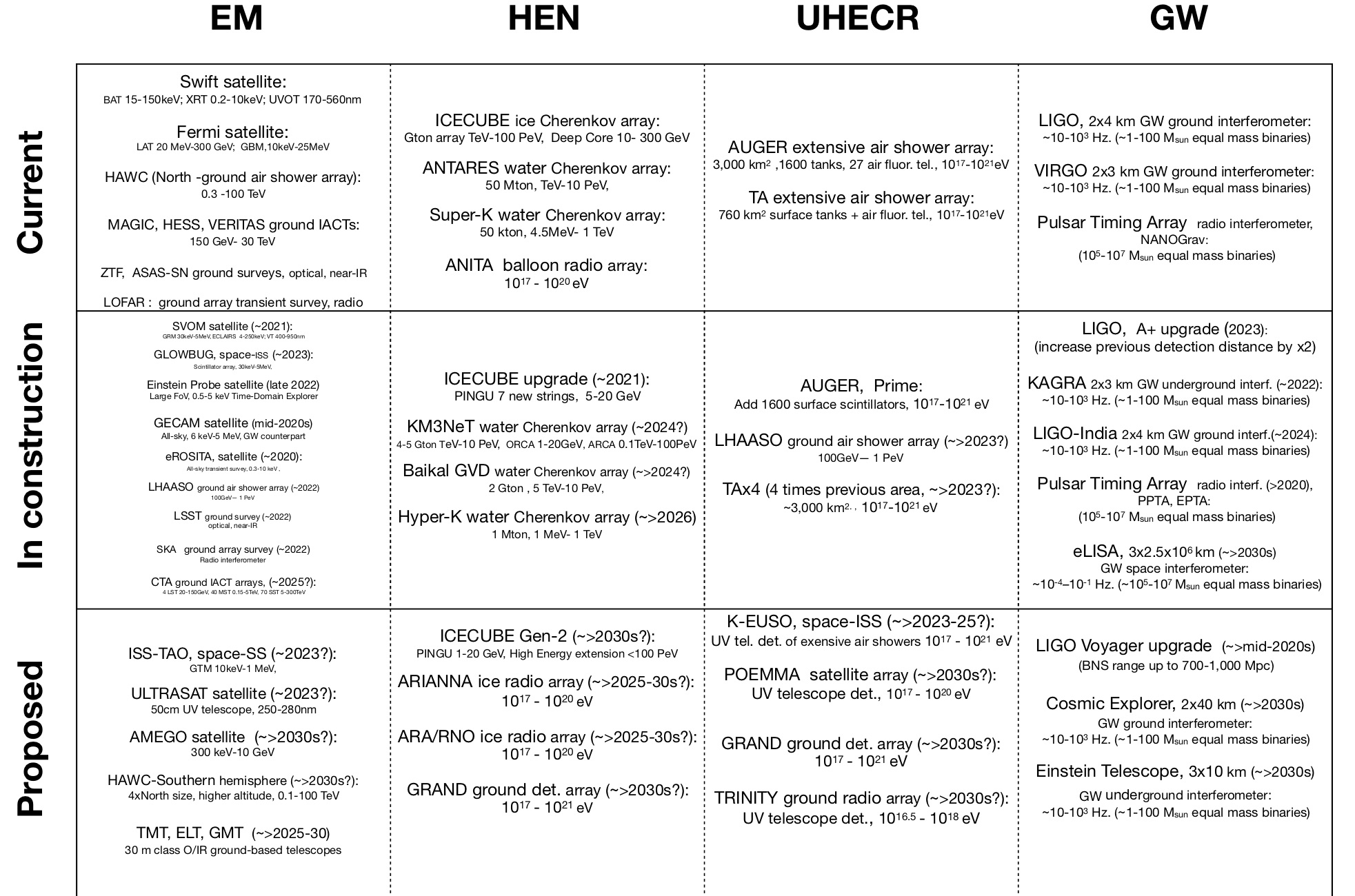}}
\end{minipage}
\caption{TABLE 1 }
\label{tab:1}
\end{figure}

\eject

\nobu {\it New Electromagnetic Detectors .-} Among the major
space-based electromagnetic facilities coming online within the next 5
to 10 years are the Chinese-French Space Variable Object Monitor
(SVOM) \cite{Dagoneau+18svom} {\bf (Fig. \ref{fig:SVOM-CTA-IC3Gen2-KM3NeT-ET}(a))}, 
designed for detecting gamma-ray, X-ray 
and optical transients, scheduled for launch in late 2022. 
{ Two other Chinese mission in preparation are {  GECAM \cite{ZHANG+19GECAM},
with an all-sky coverage of the sky aimed at detecting GW counterparts in the 
6 keV to 5 MeV energy range, scheduled for the mid-2020s}; and the time-domain 
Explorer-class Einstein Probe (EP) \cite{Yuan+15-EinsteinProbe}, with a 3600 
sq.deg. field of view sensitivity at 0.5-5 KeV, scheduled for end of 2022.} 
{ An Israeli-US mission called ULTRASAT \cite{Sagiv+14-ULTRASAT} 
has been proposed, with an ultraviolet (250--280 nm), fast slewing
($\sim$ minutes) imaging detector of 250\,deg$^2$ field of view,
which could  detect hundreds of supernovae, $\sim$10 BNS counterparts per year, 
and $\sim$100 tidal disruption events per year.}
%
{  There is a very strong case to be made for a US X-ray-\gray ~satellite for
providing real-time triggers and data, which would be critical for multi-messenger studies.}
A possible NASA mission that recently completed Phase~A
study is the ISS-TAO ``Transient Astrophysics Observer'' \cite{Yacobi+18-ISS-TAO},
on the International Space Station, with a GTM
\gray\ transient monitor and WFI wide field (350 sq.deg.) lobster-type
\xray\ imager, whose prime target would be EM counterparts of GW
sources, and which might fly by 2032.
%
%
A significant role in detecting or confirming transients of multi-messenger
importance will be played by the ZTF (Zwicky Transient Facility) \cite{Patterson+19-ZTF})
and the ASAS-SN facility \cite{Kochanek+17-ASAS-SN}.
Also in the 5 to 10 year timeframe, the multi-national Cherenkov Telescope Array (CTA) 
\cite{CTA+17science} {\bf (Fig. \ref{fig:SVOM-CTA-IC3Gen2-KM3NeT-ET}(b))}
and the Chinese Large High Altitude Air Shower Observatory (LHAASO) 
\cite{Disciascio+16-LHAASOrev} ground-based facilities will survey the sky at 
TeV-PeV gamma-ray energies, e.g.\ \cite{Montaruli19ctalhaasogam}.  
{  Both of these use the air Cherenkov technique, but while CTA includes steerable dishes 
that provide good angular localization, useful for point sources, LHAASO largely observes 
as the sky goes by, as does HAWC, which works better for extended or diffuse emission.}
A major optical/IR survey facility instrument is the Spectroscopic Survey 
Telescope (LSST) \cite{LSST+17obsstrat}, while the Square Kilometer Array (SKA) 
\cite{Mcpherson+18-SKA} will provide milliarcsecond spatial resolution images at 
radio frequencies. { Also in preparation are the 30-meter class TMT, ELT and GMT 
optical/IR ground-based telescopes \cite{Sanders+13-TMT,Varela+14-ELT,Johns+12}.}

For the early 2030s, the European Space Agency ESA is preparing a
major flagship \xray\ mission called ATHENA \cite{Nandra+11-ATHENA}, which will
trace the galaxy formation and metallicity evolution of the Universe
with its large area detectors, and can study Population III GRBs.  Due for
final ESA selection in 2022 for a launch in 2032 is the smaller but
nimbler, fast-slewing ($\sim$ minute) satellite THESEUS
\cite{Amati+18theseus}, designed to discover long GRBs at redshifts
$z\simg 9$ and seek BNS counterparts with a soft \xray\ imager, an X-
and \gray\ spectroscopic imager and an \mbox{0.7-m} class infra-red
telescope, which will also provide triggers for ATHENA.  The NASA
AMEGO satellite \cite{AMEGO+17icrc}, sensitive to \grays\ from 0.2\,MeV to
$\simg$10\,GeV and the German eROSITA \cite{Predehl+16-eROSITA} 0.3--10 keV space
detector will play important roles in in X- and \gray\ astronomy, as
well as in the EM detection of hidden neutrino sources.
{ To complement the above large facilities, it is extremely important to have
also various fast, large field-of-view robotic ground-based telescope systems
which can follow up transient candidates within seconds after an alert is triggered.}
\\

\nobu {\it High and Low Energy Neutrino Detector Improvements and Plans.-} 
High energy neutrino detector planned upgrades include the
IceCube High Energy Array and the denser PINGU sub-array, as part of an extended (10 Gtons) 
IceCube Gen2 \cite{IC3+17icrcgen2} {\bf (Fig. \ref{fig:SVOM-CTA-IC3Gen2-KM3NeT-ET}(c))}.  
In the Northern hemisphere, the completion in the Mediterranean sea of the
KM3NeT \cite{Km3net+18sens} 3 to 4 Gton EU detector is expected by
$\sim$ 2026 {\bf (Fig. \ref{fig:SVOM-CTA-IC3Gen2-KM3NeT-ET}(d))}; the error box 
improvements for muon tracks are expected to be $\siml 0.3-0.5$ deg$^2$.  
{ The relative advantages/disadvantages of ice vs. water as a Cherenkov
detector medium are that the light absorption length in ice is $\sim 100$ m
vs. $\sim 15$ m for clear ocean water; while the scattering length for 
ice is $\sim 20$ m vs. $\sim 100$ m for water. Also ocean water contains 
radioactive $^{40}$K, affecting the energy resolution signal to noise ratio.
Thus one gets relative better/worse energy resolution, and worse/better
angular resolution in ice/water, e.g. \cite{IC3+17icrcgen2,Km3net+18sens}.}
Another Gigaton water-based neutrino detector, Baikal-GVD \cite{Baikal+18-Baikal-GVD}, 
in the lake Baikal, Russia, is expected by $\sim$ 2021-22.  
Goals include determining large scale anisotropies and
individual source identifications by neutrinos alone or in tandem with
other multi-messengers. They will facilitate the use of doublets and
multiplets for source population studies, and increase the prospects
for identifying galactic sources, reliably identify tau neutrinos, and
determine the flavor composition of the high energy neutrino background.
%
%

The Hyper-Kamiokande (\mbox{Hyper-K}) \cite{Hyper-K+18instr} next generation
megaton water Cherenkov detector, operating at MeV to GeV energies, is
located in the Kamioka mine (Japan) and is scheduled to begin
construction in 2020. It will be an order of magnitude larger than its
predecessor instrument \mbox{Super-K}, where the addition of Gadolinium to
the water is providing significantly greater sensitivity. It will be
able to detect individual supernova explosions out to $\sim$4\,Mpc,
roughly one every 3 to 4 years, and in 10 to 20 years  it could
measure the relict supernova diffuse neutrino flux in the 16--30 MeV
energy range \cite{Migenda+17Hyper-K}.

At the highest energies, $10^{17}\eV$ to $10^{21}\eV$, the ANITA
balloon experiment \cite{Allison+18anita3res} will over the next several years
undergo further sensitivity improvements. On a longer timescale of
2022-2032, there is ongoing work and plans for much larger
ground-based detectors using the Askaryan radio technique for
detecting neutrinos and UHECRs in the same $10^{17}-10^{21}\eV$ range,
such as ARIANNA \cite{Barwick+17arianna} and ARA
\cite{Allison+16-ARASproto}, or a possible combination of parts of
these efforts (RNO/ARA). These are aimed at detecting the cosmogenic
neutrino background component produced by GZK UHECRs, as well as for
probing more deeply the nature of the decline of the UHECR spectrum
beyond $10^{20}\eV$. Another large-scale detector proposal aimed at this 
goal is the Chinese-led GRAND 10,000 km$^2$ array being planned for the 
2025 to 2030s \cite{GRAND+18scides}, as well as the POEMMA
\cite{POEMMA+17instr} and Trinity \cite{Otte18-trinity} projects.
\\
 
\nobu {\it UHECR Detector Improvements and Extensions.-} UHECR
facilities undergoing major upgrades include the Auger-Prime addition
of 1600 surface scintillation detectors on top of the existing water
Cherenkov tanks as well as updated electronics \cite{Auger+17-ICRC}; and the
Telescope Array upgrade to the TAx4 configuration four-fold surface
enlargement \cite{Sagawa+15-TAx4}. In the next 5-10 years the planned
K-EUSO experiment \cite{Casolino+17-K-EUSO} on the International Space Station
(ISS) could achieve uniform exposure across the Southern and Northern
hemispheres of $4\times 10^4\,\km^2\,\sr\,\yr$ per year, an order
of magnitude larger than Auger or TAx4. Radio observations with LOFAR 
(Low Frequency Radio Array) \cite{Winchen+19-LOFARcr} may also contribute 
substantially to an understanding of UHECRs.
In the lower energy range of $10^{12}$ to $>$10$^{15}$\,eV the ISS-CREAM
\cite{Seo+14-ISSCREAM} on the International Space Station, building on
earlier work of the Cosmic Ray Energetics And Mass (CREAM) balloon flights, 
{  as well as the future Chinese mission HERD \cite{Zhang+14HERD}},
could greatly increase knowledge about the spectrum and
compositions of CR nuclei with charges in the $Z=1-26$ range.
In the next 10 to 15 years the planned POEMMA \cite{POEMMA+17instr} is
expected to achieve an increase in the exposure by 100$\times$, while
the planned FAST ground array \cite{FAST+17uhecr} would provide 10x
the exposure with high quality events. These advances will address the
chemical composition and anisotropy issues of UHECR, the 
{  interpretation of the surmised three CR components making up 
the entire spectrum (e.g. see Fig. 4 of \cite{Gaisser+13crsp}) },
the maximum energy of galactic CRs, and will probe
in much more detail the nature of the spectral cutoff, the transition from
galactic to extragalactic components, the strength of the galactic and
intergalactic magnetic fields, etc.  \\

\begin{figure}[H]
\centerline{\includegraphics[width=6.0in,height=1.7in]{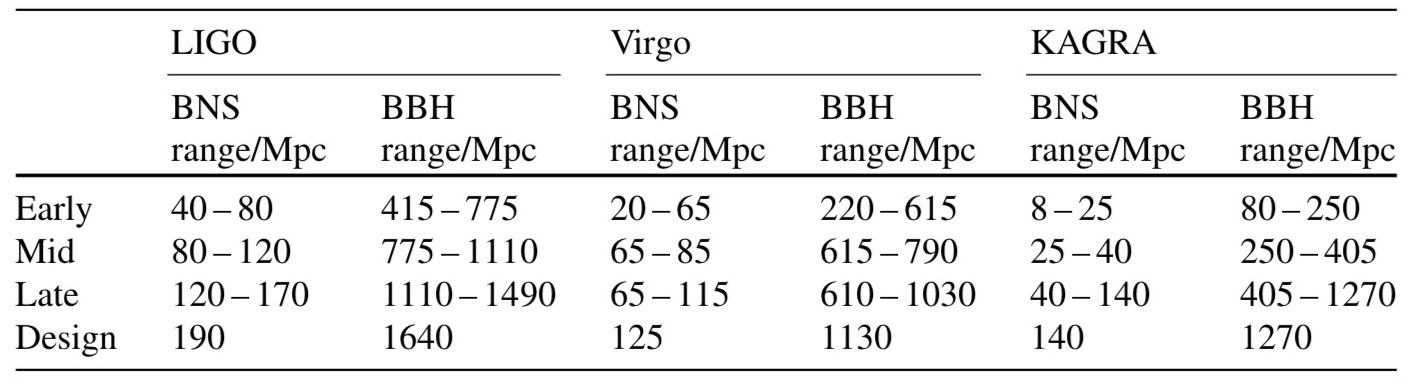}}
\caption{\footnotesize
TABLE 2. Plausible target detector sensitivities, giving the average detection distance (range)
at which a $2\times 1.4\msun$ BNS and $2\times 30\msun$ BBH may be detected with
LIGO, VIRGO and KAGRA \cite{Abbott+18gwlivingrev}.
}
\label{tab:LVKrange}
\end{figure}

\nobu {\it Gravitational Wave Detectors Improvements \& Planned New
  Facilities.-} The upgrades of LIGO and VIRGO are continuing
\cite{Abbott+18gwlivingrev}, and this will reduce the $90\%$ median
credibility error box size for source identifications down to $120-180$
deg$^2$ (with 12-21\% with $\leq 20$ deg$^2$) by the 2019+ O3 run. The
Japanese KAGRA detector being built in the Kamioka mine in Japan is
expected to reach a sensitivity comparable to aLIGO/aVIRGO by 2024. A
new LIGO observatory is under construction in India to house the third
detector (LIGO-India). With these additional facilities, the expected
median 90\% localization error box sizes will be 9-12 deg$^2$. The
number of expected binary black hole and neutron star detections and
the limiting distances are shown in Table 2 {\bf (Fig. \ref{tab:LVKrange})}.

Among the next generation of ground-based GW detectors planned for the
2020-30s \cite{Abbott+17gwnextgen}, {  in the US, the ``A+'' upgrade 
to the LIGO facilities has been approved, which should provide a further factor 
of two increase in detection range beyond the detectors' advanced sensitivity.
{  A further upgrade, termed LIGO Voyager, would have a new detector operating in 
the existing LIGO, with cryogenic mirrors in the existing LIGO vacuum envelope. 
This could bring a further factor of 3 increase in the BNS range (to 1100 Mpc),
with a low frequency cut-off down to 10 Hz.}
In the EU, the planned underground Einstein Telescope \cite{Sathyaprakash+12-EINSTEIN}
{\bf (Fig. \ref{fig:SVOM-CTA-IC3Gen2-KM3NeT-ET}(e))},
with three arms of 10 km length, will be able to measure the GW polarization } 
of BBHs and BNS sources from distances 3 to 10 times farther than with the
current design sensitivity of the LIGO/VIRGO designs. 
Further in the future, the Cosmic Explorer ground-based observatory 
\cite{Abbott+17-COSMICEXPLORER-GWNextGen} would use 40~km arms to achieve 
a further order of magnitude improvement in sensitivity over the 10 to 10$^4$ Hz 
frequency range and detect compact binary in-spirals throughout our Hubble volume.
%
%
%

In order to study the merger of the much larger ($10^6$ to
$10^8\msun$) ``supermassive'' black holes located at the center of
galaxies, the EU is planning a large space-based GW detector called
eLISA \cite{Klein+16-eLISA}, consisting of an interferometer using three
small satellites in Solar orbit.  A different technique for the
exploration of supermassive BH mergers is provided by the Pulsar Timing
Arrays (PTAs) \cite{Schutz+16ptambh}, such as NANOGrav, PPTA, and
EPTA, e.g.  \cite{Hobbs+17ptagw,Arzoumanian+18nanograv11}.  This
technique relies on measuring the time delays in the EM radio signals
from distant pulsars caused by the space-time variations induced by
the GW field of merging BH binaries, and is expected to yield the
first stochastic (population-integrated) detections in the near future.  \\

\nobu {\it Exploiting the Multi-Messenger Synergy.-} The Astrophysical
Multi-messenger Observatory Network (AMON) \cite{Keivani+17AMON} is a
multi-institution consortium which has signed MoUs with a number of
observatories using different messengers. { One major purpose is to
combine disparate rare signals appearing in coincidence, so that even sub-threshold 
detections in one messenger, when combined with other sub-threshold signals 
in other messengers, can yield a reliable above-threshold detection.  
The other purpose is to rapidly distribute interesting transient alerts
arising from any observatory to all the other observatories and the community,
to facilitate rapid follow-up.}
The architecture of AMON consists of a
central hub with radial spokes connecting to individual observatories,
from which it receives individual sub-threshold (and also above threshold)
alerts, which are then subjected to analysis and/or redistributed to
other observatories. This greatly increases the speed and
effectiveness of reaction to a trigger, compared to the large number
of traditional individual observatory-to-observatory connections. 
Observatories that have signed up to the AMON network
include, so far, ANTARES, Auger, \fermi\ LAT, \fermi\ GBM, HAWC,
IceCube, Swift-BAT, VERITAS, and others.  Live data streams from
IceCube, HAWC, \fermi\ and \swift\ are being received by AMON,
triggering alerts as in the \mbox{TXS~0506+056} blazar neutrino and
gamma-ray flare discovery \cite{Keivani+18txs0506,Ayala+19-AMONmulti},
and other coincident sub-threshold analyses are being carried out,
e.g.\ for LIGO + \swift\ BAT, HAWC, and others.   The unique and 
most promising aspect of AMON is its massive emphasis on leveraging
multiple live sub-threshold alerts.  In addition, sub-threshold
analyses are also being carried out using archival data from different
individual observatories, e.g.\ \cite{Turley+18nugamsearch}. Other
groups are also developing algorithms and strategies for
multi-messenger searches, e.g. \cite{Countryman+19llama}.  
\\

\nobu {\it Theory and Simulations.-} The high-quality data provided by
the facilities outlined above will only yield fruit insofar as it
is thoroughly analyzed and interpreted through realistic source models
satisfying state-of-the-art physics (while keeping in mind the ultimate
possibility of beyond the standard model physics).  Such models must
be considered at three levels. The basic level is based on an overall
conceptual picture, using analytical or semi-analytical source models,
with their constitution, dynamics and multi-messenger radiation or
micro-physical properties, e.g.\ \cite{Murase+13pev,Ahlers+14pevnugal,Murase+18txs0506}, see
\cite{Ahlers+18ic3nurev,Schutz18gwrev,Murase+16uhenurev,Meszaros17arnpnurev,Abbott+18gwlivingrev}
for reviews.
The second level involves detailed numerical simulations of the dynamics of the formation of the sources 
and their evolution leading to the state at which the various types of multi-messenger emission occurs, 
e.g. \cite{Shibata+17grhydrobns,Easter+18gwbns,Parsotan+18grbphotsim,Vaneerten+18grbagrev}.
The third level involves detailed calculations and simulations of the radiation physics, using 
large-scale numerical codes to describe the emission of multi-messengers, followed by their possible 
changes during propagation from the source to the observer, and their detailed effects on particular 
types of detectors, e.g. \cite{Fang+18crnuagnjet,Senno+16hidden,Keivani+17AMON,Hotokezaka+18nsmerg,
Biehl+18crnugrb,Alves+19cosmonu}.
For low source numbers or low signal rates, diffuse backgrounds must be calculated using simulated
source signals convolved over cosmological luminosity and redshift distributions, 
e.g. \cite{Alves+19cosmonu}.

All three of these types of calculations will have to be considerably expanded and refined to address 
and exploit the potential of the much more detailed data expected from the above new facilities.
Even for the semi-analytical studies, the increasing sensitivity and range of the detectors will
make it necessary to make use of farther and fainter reaching source luminosity functions, more
extensive redshift and mass distributions, intervening plasma and radiation field spectral densities, etc.
The source formation and dynamics studies, which are increasingly incorporating general relativistic
and magnetohydrodynamic (MHD) effects, will need to be extended to the 3-dimensional regime much more
commonly than before, and the use of adaptive mesh and shock capture methods will have to be exploited
and developed further. There will be an increasing need for a better cross-calibration of the various
Monte Carlo methods used in simulating high energy particle interactions and cascades, incorporating 
the errors due to theoretical uncertainties, especially in the extrapolation to energies beyond those of 
accelerator data. As another example, the atomic and nuclear physics of very heavy elements is poorly
understood and sparsely studied in the lab, yet to reliably elucidate the sources of r-process elements 
in the Universe (lately ascribed largely to BNS mergers), the error estimates arising from these
theoretical  and experimental lab uncertainties will need to be quantified and taken into account.
{  We must emphasize that nuclear and atomic physics lab experiments that can shed light on the 
r-process details are crucial for our understanding of this important phenomenon.}
%


\section{\large Conclusions and Perspectives}
\label{sec:conclusion}

Some of the most important questions that will be addressed in the
next 5 to 10 years with upgraded GW detectors such as LIGO, VIRGO and
KAGRA, as they improve sensitivity at frequencies above 0.1\,kHz, are
to explore in detail the lower mass range of binary black hole
mergers, to test whether the final outcome of neutron star mergers is
a massive neutron star or a black hole, to probe the final ringdown of
space-time around a newly formed black hole, to determine the maximum
mass of neutron stars, to test whether General Relativity remains valid
under extreme density and pressure conditions, and to explore the nature
of the central engine of GRBs.  As they push towards lower frequencies
below 10 Hz it will be possible to probe intermediate mass BHs of 100
to 500 solar masses, important for understanding how the most massive
BHs at the center of galaxies form. To get more reliable results, larger 
interferometer arm lengths such as the $\sim$ 10 km of the Einstein or 
Cosmic Explorer experiments will be needed. On the 10-20 year timescale, 
even longer arms, such as the 2.5 million km in the space-based eLISA
interferometer, will measure frequencies $\sim 0.1$ Hz which can probe
the merger of $\simg 10^6\msun$ black hole mergers, important for
understanding the growth of galaxies and clusters of galaxies in the
Universe, and the existence of a primordial GW background left over
from the inflationary era of the early Bing Bang.

The neutrino detector upgrades such as PINGU in IceCube and ORCA in
KM3NeT will get better statistics in the 1-10 GeV energy range to
probe fundamental neutrino physics issues such as the hierarchy of the
mass ordering, as well as the mixing angles between flavors,
constraining the neutrino masses and testing the existence of sterile
neutrinos. These issues are likely to be resolved in the next 5-10
years { with the help of these and other detectors}, 
while the next generation IceCube Gen-2 is likely to identify the sources 
of the origin of high energy neutrino, e.g.\ \cite{Murase+16uhenurev}. 
In the Northern hemisphere, the 3 to 4\,Gton KM3NeT detector will be 
able to observe HENs from our galactic center, where most of the 
(so far undetected) galactic neutrino sources reside.
The upgraded ANITA balloons and, if funded, the ARIANNA and RNO/ARA
experiments in Antarctica will 
{ make progress towards detecting the $10^{17}-10^{19}$ eV cosmogenic 
neutrinos produced by UHECR protons interacting with the cosmic microwave background, 
or by UHECR nuclei interacting with the diffuse starlight. These experiments will 
also complement IceCube-Gen2 in the pursuit of the identification of tau-neutrinos.}
More reliable determinations may need to wait for larger experiments such as 
GRAND and POEMMA, which will also address the chemical composition of the
highest energy UHECRs.
The next Galactic supernova should be an ideal and important event for
multi-messenger astrophysics, which can be exquisitely studied by the
Hyper-K \cite{Hyper-K+18instr} and JUNO \cite{Lu+16junosnnu}
experiments to address fundamental issues of neutrino oscillation and
supernova physics, e.g.\ \cite{Beacom10sndifnu,Tamborra+18ccsnnu}.

The next upgrades of the Auger and TA UHECR arrays are expected to
answer the chemical composition question independently of the answers
obtained through the above neutrino detectors, providing a much needed
consistency check.  When the TAX4 array is completed its area will be
comparable to Auger's, and being in the Northern hemisphere while
Auger is in the Southern, will prove or disprove possible anisotropies
of arrival, also testing whether the UHECR production is dominated by
a few nearby sources or more numerous distant ones.  Together, both of
them will probe the details of the properties of hadronic interactions
at energies three orders of magnitude higher than what is achievable in
laboratory accelerators.  These questions can be more thoroughly
investigated from space with K-EUSO and POEMMA in the 10-15 years.

Future all-sky monitors such as the Large Spectroscopic Survey
Telescope (LSST) in the optical and the Square Kilometer Array (SKA)
in the radio, as well as the \fermi, \swift\ and expected (2022) SVOM
satellites at keV to GeV energies will provide rapid EM triggers and
accurate sky localization as well as follow-up capability for studying
cataclysmic events such as BH or NS mergers, supernovae, gamma-ray
bursts, AGN flares, etc., where CRs, neutrinos and GWs are also
expected in varying amounts. The strength and mix of these different
messengers is model-dependent, and multiple triggers in different
messengers, as well as model development and extensive simulations,
will be the key for understanding the physics of these energetic
sources.

The key to fully exploit the power of these new facilities is the
multi-messenger approach, due to the complementary advantages and
limitations of the different messenger particles. Cosmic rays provide
unique information about particle accelerators well beyond terrestrial
laboratory capabilities, and about source magnetic fields and total
energetics, but they do not reach us from beyond $\sim$100\,Mpc and have at
best poor angular resolution.  Neutrinos reach us from the most
distant reaches of the Universe, and probe the inner, denser regions
of the most energetic and cataclysmic events, and ultra-high energy 
neutrinos, being intimately linked to UHECRs, can provide unique clues 
as to how the latter reach their enormous energies.  Gravitational wave
observations probe the most compact regions of high energy sources; the
GW wave strain amplitude at earth is directly proportional to the source
compactness, measured in terms of $GM/c^2R$, and the GW luminosity goes 
as the fifth power of the compactness. GWs provide excellent information 
about central object masses, angular momenta, and distances, and they will 
eventually be detected from the farthest distances and earliest epochs in the
Universe, but they have modest angular angular resolution at best, and do
not probe the bulk of the stellar or baryonic mass of their sources.
Electromagnetic waves provide excellent angular resolution, velocity
and redshift determination capabilities, but the opacity of matter
prevents them from probing the inner, denser regions of astronomical
sources, while at higher gamma-ray frequencies they cannot reach us
directly from the much grater distances probed by neutrinos or
gravitational waves.  By using several messengers in conjunction,
astronomers and physicists can hedge the foibles of each against the
advantages of the others, forging them into a formidable toolkit
for probing the highest energies and densest, most violent corners of
the Universe, and in this fashion, putting our physical theories of
the universe to their most extreme and exacting possible tests.

\section{Explanatory Boxes}
\label{sec:boxes}


{\bf BOX 1: Neutrino Production and Oscillations}\\
\noindent
Cosmic ray protons $p_{cr}$ interacting with target protons $p_t$ and target photons $\gamma_t$ 
lead to a reduced energy proton $p$ or neutron $n$, and to intermediate charged or neutral 
short-lived unstable particles, such as pions $\pi^{\pm,0}$, muons $\mu^\pm$, neutrons $n$ 
and (at higher energies) Kaons $K^\pm$, whose decay results in neutrinos $\nu_i$ of different 
flavors $i$, $\gamma$-rays and electrons or positrons $e^\pm$. 
%
{  CR nucleons, on the other hand, are primarily subject to spallation against target protons and
photo-disintegration against target photons, both resulting in smaller nucleons including protons, 
the latter subsequently undergoing the same above mentioned interactions.}
\\

\beq
p_{cr} + p_t/\gamma_t \to    p/n + \pi^\pm + \pi^0 + K^\pm + \cdots
\enq
\bea
                  \pi^+ \to & \mu^+ + \numu , ~~~~~~~~~~~~~\cr
                            & \mu^+ \to e^+ + \nue + \barnumu \cr
                  \pi^- \to & \mu^- + \barnumu , ~~~~~~~~~~~~~\cr
                            & \mu^- \to e^- + \barnue + \numu
\ena
\bea
                   \pi^0 ~~\to & \gamma + \gamma \cr
                   K^+ /K^- \to & \mu^+ / \mu^- + \numu /\barnumu \cr
                   n     \to & p + e^- +\barnue
\ena
%

\noindent
The primary CR proton's mean relative energy loss per interaction, called the inelasticity, is
$\kappa_{pp}\sim 0.5$ for $pp$ (or $pn$) and $\kappa_{p\gamma}\sim 0.2$ for $p\gamma$ interactions.
For $p_{cr}$ interactions with target
protons (or target neutrons) the mean ratio of charged to neutral pion secondaries is $r_{\pm/0}\sim 2$, 
and for interactions with target photons it is $r_{\pm/0}\sim 1$. The electrons and positrons $e^\pm$
quickly lose their energy via synchrotron or inverse Compton scattering, resulting in further $\gamma$-rays,
so the final result of the $p_{cr}+ p_t/\gamma_t$ interactions are high energy neutrinos $\nu_i$ and 
$\gamma$-rays.  The mean energy of the resulting neutrinos and $\gamma$-rays is $\sim 0.05$ and $\sim 0.1$ 
of the initial CR proton's energy.

Once neutrinos of any flavor are produced, during their travel from the source to the observer 
the neutrinos of any flavor can change into neutrinos of any of the three flavors, in the so-called 
neutrino oscillation phenomenon. As a result, independently of the ratio of neutrino flavors 
initially produced a the source where the CR interactions took place, after traveling over astronomical
distances typically all three neutrino flavors are present at the observer.
{ For the neutrino sources considered here, the observable neutrino flavors are expected to be 
oscillation-averages due to the large source distances and the finite energy resolution of 
the neutrino observatories.}
\\

\eject


{\bf BOX 2: Multi-messengers and their inter-relation~ }
\begin{figure}[H]
\begin{minipage}{0.56\textwidth}
\centerline{\includegraphics[width=4in]{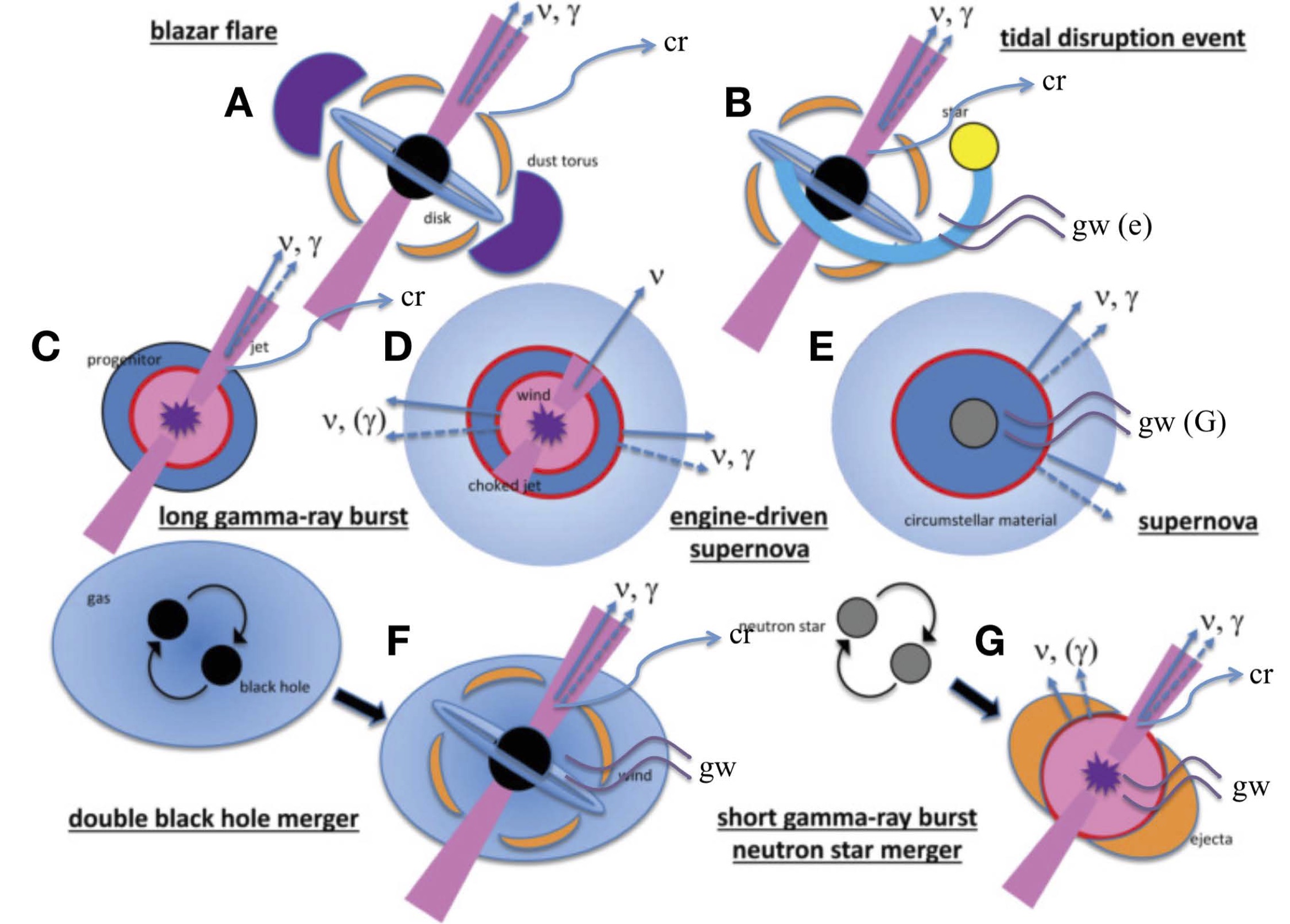} }
\end{minipage}
\hspace{9mm}
\begin{minipage}[t]{0.32\textwidth}
\vspace*{-1.3in}
\caption{\footnotesize
Multiple messenger particles {  possibly emanating from 
(A) a blazar flare; 
(B) a tidal disruption event (gravitational waves detectable by eLISA for some events); 
(C) a long gamma-ray burst;
(D) an engine-driven supernova, or 
(D) a supernova (gravitational waves detectable for Galactic events); 
(F) a double black hole merger, or 
(G) a double neutron star merger leading to a short gamma-ray burst.} }
\label{fig:multi-transient-fig}
\end{minipage}
\end{figure}
A multi-messenger source might emit two, three, or even all four
different types of messengers.
{  
For instance, panel (G) of Fig.\ref{fig:multi-transient-fig} shows a binary
neutron star merger such as the GW/GRB 170817 event, from which two
types of multi-messengers, gravitational waves (GW) and photons ($\gamma$), 
were observed \cite{LIGO+17gw170817disc,Troja+17gw170817xr,
Kasliwal+17gw170817concord}. Such sources may also emit high energy
neutrinos (HENs) and cosmic rays (CRs) e.g.
\cite{Kimura+17sgrbnu,Kimura+18transejnu,Kimura+18bnscr}, although for
this particular source theories predict fluxes too low for current
detectors; if so, even closer events or next-generation HEN facilities
will be required to observe HEN from these sources. 
Another panel, (B), shows a tidal disruption event (TDE) of a star} 
by a massive black hole; in this case shocks in the disrupted gas can
accelerate particles and lead to CRs and HENs,
e.g.\ \cite{Guepin+17tdecrnu,Biehl+17tdecrnu,Senno+17tde,Wang+16tdecrnu}. 
TDEs involving white dwarf stars and $\sim$1000\,\msun\ black holes lead to strong
low-frequency ($\sim$1\,mHz) gravitational wave emission that could be
observed by the forthcoming eLISA mission. 

{  A solitary supermassive black hole with a jet may emit \grays, HEN, 
and cosmic rays (panel B)}, 
as we suspect occurred during the 2017 flaring episode of the BL~Lac-type 
blazar \mbox{TXS~0506+056} \cite{Magic+17blazarnu,Keivani+18txs0506,NuSTAR+17blazarnu,
  IC3multi+18txs0506,Gao+19txs0506,Cerruti+19txs0506}. 
Here and in related sources, the coproduction of CRs, HEN, and
high-energy \grays\ is anticipated, as the physics of these three
messengers are closely connected -- high-energy particle acceleration
and shocks lead to the interaction of highly-relativistic protons (or
nuclei) with ambient gas or intense radiation fields, resulting in
neutrinos, \grays, and $e^\pm$. 
For single objects, even those of extreme mass and undergoing
substantial accretion, relatively weak gravitational wave emission is
expected as the time-varying quadrupole moment (which requires the
breaking of azimuthal symmetry) in these cases are thought to be
minimal. {  The sole exception would be a Galactic engine-driven 
supernova or a Galactic supernova (panels D and E), which would be}
sufficiently nearby that detection of coherent or incoherent
gravitational waves by current and future ground-based detectors is
anticipated. {  As far as thermal ($\sim 10$ MeV) neutrino detection
from such supernovae, IceCube is well matched, having for such emission
also a roughly galactic reach.
A challenge for theory is to predict the amplitude and 
spectrum of GW and neutrinos from different types of supernovae. }

Strong GW emissions have been observed from the mergers of compact
binary systems, either from {  two merging stellar-mass black holes (panel F)
\cite{LIGO+18bbhGWTC-1}, two merging neutron stars (panel G)}
\cite{LIGO+17gw170817disc}, or (in the future) BH-NS mergers, because
the final in-spiral to coalescence yields a strong gravitational wave
signal in the ``sweet spot'' frequency range for ground-based
gravitational wave detectors. In the case of 30\,\msun\ +
30\,\msun\ black hole binary systems, such coalescence events can
already be observed out to $\sim$500\,Mpc distances
\cite{Abbott+18gwlivingrev}.
However, in the case of BH-BH mergers little EM flux is expected,
because the ambient matter density (protons, electrons) in the
vicinity of the binary, at the time of the merger, is typically very
low.
A key exception is the accreting supermassive black holes at the
centers of massive galaxies, which are expected to merge in the wake
of the coalescence of their component galaxies. These SMBH mergers are
key targets for the eLISA mission, and may well exhibit accompanying
EM, CR, and HEN emission \cite{Klein+16eLISA-MBH}.
\\


\noind
{\it Acknowledgments:}  We are grateful to Stephane Coutu, Douglas Cowen, Miguel Mostaf\'a and
Bangalore Sathyaprakash and the referees for useful discussions and comments.


\bibliographystyle{revtex}
\footnotesize


\input{nat-arx2.bbl}

\end{document}

%% file: macos.tex
\def\mathnew{\mathsurround=0pt}
\def\simov#1#2{\lower .5pt\vbox{\baselineskip0pt \lineskip-.5pt
       \ialign{$\mathnew#1\hfil##\hfil$\crcr#2\crcr\sim\crcr}}}
\def\simg{\mathrel{\mathpalette\simov >}}
\def\siml{\mathrel{\mathpalette\simov <}}

\def\msun{\mbox{$M_\odot$}}

\def\swift{\textit{Swift}}
\def\fermi{\textit{Fermi}}

\def\noind{\noindent}

\def\nobu{\noindent$\bullet$\,\,}

\def\beq{\begin{equation}}
\def\enq{\end{equation}}
\def\bea{\begin{eqnarray}}
\def\ena{\end{eqnarray}}
\def\bec{\begin{center}}
\def\enc{\end{center}}

\def\blist{\begin{list}{$\bullet$}{\itemsep 0.0in \parsep 0.0in}}
\def\elist{\end{list}}
\def\bitem{\begin{list}{\arabic{enumi}.}{\usecounter{enumi} \itemsep 0.0in \parsep 0.0in}}
\def\eitem{\end{list}}

\def\PeV{\hbox{PeV}}
\def\TeV{\hbox{~TeV}}

\def\eV{\hbox{~eV}}
\def\xray{\mbox{X-ray}}
\def\xrays{\mbox{X-rays}}
\def\gray{\mbox{gamma-ray}}
\def\grays{\mbox{gamma-rays}}

\def\part{\partial}

\def\nue{\nu_e}
\def\barnue{{\bar \nu_e}} 
\def\numu{\nu_\mu}
\def\barnumu{{\bar \nu}_\mu}

\def\km{{\rm km}}

\def\Mpc{{\rm Mpc}}

\def\sr{{\rm sr}}

\def\yr{{\rm yr}}

\def\m-pl{m_{Pl}}

\def\h75{h_{75}}
\def\Omh75{\Omega h^2_{75}}
\def\Omh70{\Omega h^2_{70}}

\def\fun#1#2{\lower3.6pt\vbox{\baselineskip0pt\lineskip.9pt
  \ialign{$\mathsurround=0pt#1\hfil##\hfil$\crcr#2\crcr\sim\crcr}}}

\def\jcap{Jour. Cosmology and Astro-Particle Phys.\,}
\def\mnras{M.N.R.A.S.\,}

\def\apj{Astrophys.J.\,}
\def\apjl{Astrophys.J.Lett.\,}

\def\aj{Astron.J.\,}
\def\nat{Nature\,}

\def\prl{Phys. Rev. Lett.\,}

\def\prd{Phys.Rev.D\,}
\def\araa{Annu.Rev.Astron.Astrophys.\,}
\def\aap{Astron.Astrophys.\,}